\documentclass[twocolumn]{revtex4}

\usepackage{xspace}
\usepackage{pdfpages}

\newcommand{\um}{\,$\mu$m\xspace}

\newcommand{\be}{\begin{eqnarray}}
\newcommand{\ee}{\end{eqnarray}}

\usepackage{latexsym}
\usepackage{graphics}
\usepackage{amssymb}
\usepackage{multirow}
\usepackage{textcomp}

\begin{document}


\title{An ion trap built with photonic crystal fibre technology}



\author{F.~Lindenfelser}
\email[]{friederl@phys.ethz.ch}
\affiliation{Institute for Quantum Electronics, ETH Z\"urich, Otto-Stern Weg 1, 8093 Z\"urich, Switzerland}

\author{B.~Keitch}
\altaffiliation{Present address: Department of Physics, University of Oxford, Clarendon Laboratory, Parks Road, Oxford OX1 3PU, U.K.}
\affiliation{Institute for Quantum Electronics, ETH Z\"urich, Otto-Stern Weg 1, 8093 Z\"urich, Switzerland}

\author{D.~Kienzler}
\affiliation{Institute for Quantum Electronics, ETH Z\"urich, Otto-Stern Weg 1, 8093 Z\"urich, Switzerland}

\author{D.~Bykov}
\affiliation{Max Planck Institute for the Science of Light, Guenther-Scharowsky-Str. 1/Bldg. 24, 91058 Erlangen, Germany}

\author{P.~Uebel}
\affiliation{Max Planck Institute for the Science of Light, Guenther-Scharowsky-Str. 1/Bldg. 24, 91058 Erlangen, Germany}

\author{M.~A.~Schmidt}
\affiliation{Leibniz Institute of Photonic Technology e.V.,
Albert-Einstein-Stra{\ss}e 9, 07745 Jena, Germany}
\affiliation{Otto-Schott-Institute for Material Research,
Fraunhoferstra{\ss}e 6, 07743 Jena, Germany}

\author{P.~St.J.~Russell}
\affiliation{Max Planck Institute for the Science of Light, Guenther-Scharowsky-Str. 1/Bldg. 24, 91058 Erlangen, Germany}

\author{J.~P.~Home}
\email[]{jhome@phys.ethz.ch}
\affiliation{Institute for Quantum Electronics, ETH Z\"urich, Otto-Stern Weg 1, 8093 Z\"urich, Switzerland}


\date{\today}

\begin{abstract}

We demonstrate a surface-electrode ion trap fabricated using techniques transferred from the manufacture of photonic-crystal fibres. This provides a relatively straightforward route for realizing traps with an electrode structure on the 100 micron scale with high optical access. We demonstrate the basic functionality of the trap by cooling a single ion to the quantum ground state, allowing us to measure a heating rate from the ground state of 787$\pm$24\,quanta/s. Variation of the fabrication procedure used here may provide access to traps in this geometry with trap scales between 100\um and 10\um.

\end{abstract}

\pacs{}

\maketitle 

\section{Introduction}
Trapped atomic ions are among the best experimental systems for quantum computation, quantum simulation and atomic frequency standards~\cite{03Leibfried2}. Ion traps in use today span from those fabricated using precision machining to microfabricated structures formed using photo-lithographic techniques. The latter offer potential advantages in the context of minituarization due to high precision of manufacture, and can potentially be used for more complex, larger scale systems following techniques transferred from the microelectronics industry ~\cite{08Seidelin,14Mehta}. Creating an interface  between trapped ions and optical fields is a key component of proposals for scalable quantum computation and communication~\cite{07Moehring2,13Monroe,14Nickerson}. One approach has been to integrate optical fibres into a dedicated trap structure~\cite{10VanDevender,11Kim,14Clarck,13Takahashi,13Brandstaetter,13Steiner}. We take the opposite route and present an ion trap which takes advantages of the early steps in the production of photonic-crystal-fibres (PCFs) to realize a surface electrode trap. This provides a simple fabrication process for traps at the 100~micron scale. Extensions of our technique may allow for traps with integrated PCFs or trap arrays suitable for quantum simulation of 2-dimensional lattices~\cite{00Cirac,12Schneider2}.

\begin{figure}
  \includegraphics{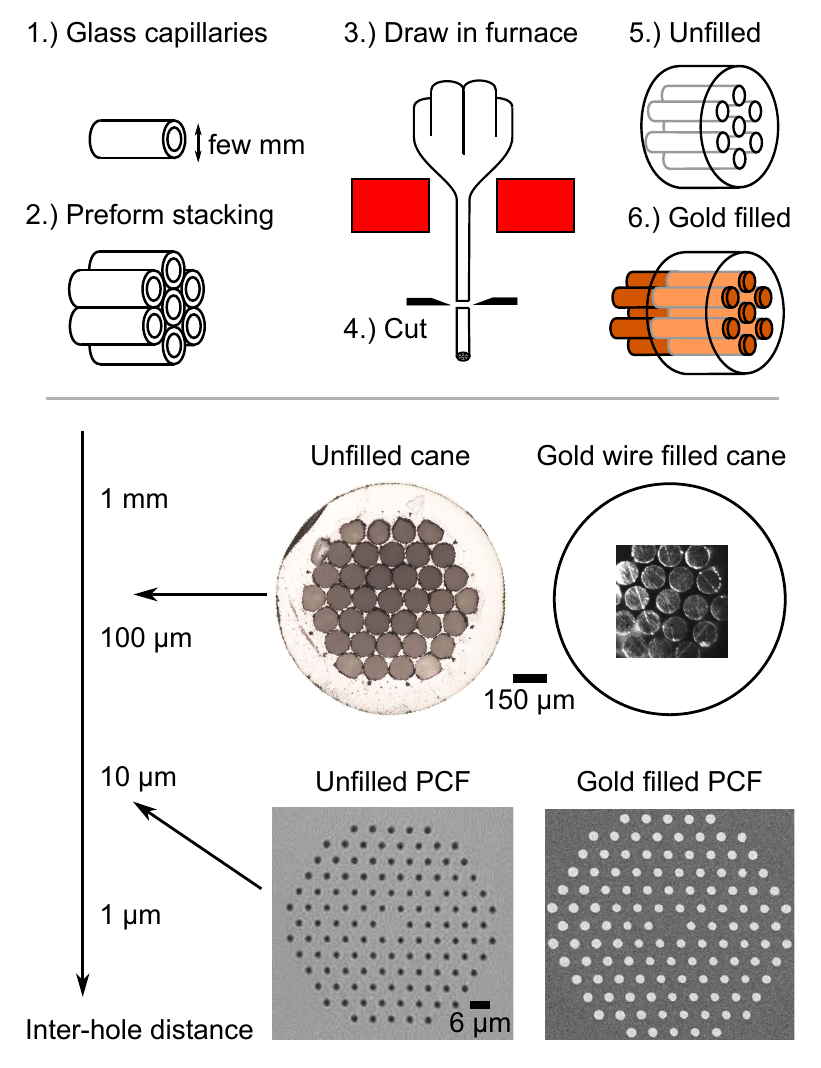}
\caption{
Production process for gold filled PCF (steps 1. to 6.) a range of possible sizes (bottom left), and two realizations, both empty and gold filled. The gold filled version on the larger scale shows the trap used in this paper, as seen through the ion imaging optics.
}
\label{fig:drawFill}
\end{figure}

\section{Gold filled PCF technology}

A solid-core PCF is a microstructured optical fiber consisting of a regular array of hollow channels (forming the cladding) surrounding a silica core (i.e. a defect channel) which run along the entire length of the fibre. An image of the cross section of an exemplary PCF is shown in the lower left corner of figure\,\ref{fig:drawFill}. The channels lower the effective index of the cladding with the effect that light is guided in a core mode by modified total internal reflection~\cite{97Birks}. Other applications of solid-core PCFs include supercontinuum generation, nonlinear optics and mode filtering~\cite{03Russell}.

One prominent technique to fabricate PCFs is stack-and-draw as illustrated in figure\,\ref{fig:drawFill}. A preform is obtained by stacking silica capillaries in the desired geometry. A standard drawing tower is used to make a fibre. There the preform is fixed at one end, with the other end heated to around 1950\,$^{\circ }$C in a furnace which fuses the capillaries together. At the same time the structure is allowed to fall under its own weight along the axis of the capillaries and the part that comes out of the furnace is pulled using a tractor system. As the glass is drawn, the cross-section is reduced. This drawing results in a cane of $\sim$1\,mm outer diameter with an array of holes parallel to the axis. Further drawing brings the structure down to the size of a fibre of $\sim$100\um diameter. By changing the speed at which the structure is fed into the furnace and pulled out of the furnace the scaling of the diameter can be adjusted. The fabrication technique in use allows unprecedented control of the arrangement and size of the hollow channels in a glass matrix. Inter-hole spacing of 0.5\um to 1\,mm and hole diameters down to several 10s of nm can be achieved.

In the past years several techniques for infiltrating the hollow channels of PCFs with gold were developed\,\cite{11Lee1,13Uebel}. At the cane stage holes of $\sim100$\um size can be manually filled with gold wires. Further drawing then leads to micro- or nano-meter size gold wires inside the fibre. More complex arrays of nanowires can be produced by pressure-assisted melt filling\,\cite{11Lee1}. These are used to investigate surface plasmon polaritons on coupled waveguide systems~\cite{08Schmidt,12Uebel,13Uebel}.

\section{PCF technology based ion trap fabrication}

The ability to connect gold wires through an insulating structure to form an array of electrodes in a surface plane has been a significant goal of ion trap fabrication research in recent years, due to the desire to realize isolated electrodes which offer advantages for multi-zone trap arrays and 2-dimensional lattices\,\cite{10Amini,10Kumph}. With the advent of metal filled PCF, one route is to fill a fibre structure with gold, and use the gold wires protruding from one end as electrodes to trap an ion at the tip of the structure.

\begin{figure}
\includegraphics{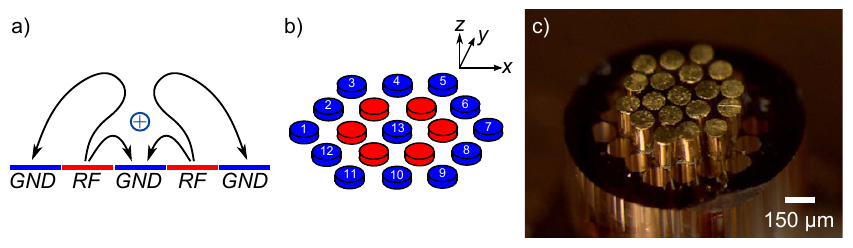}
\caption{
Point trap created with honeycomb electrode pattern. a) Sketch of quadrupole potential generated by alternating grounded and RF electrodes. The position in which ions can be trapped is indicated by the cross. b) planar structure of electrodes that forms a quadrupole potential as sketched in a). Non RF electrodes are numbered and a coordinate system is introduced. c) Photo of a gold wire filled PCF cane that exhibits this structure.
}
\label{fig:RingTrap}
\end{figure}

In order to test this idea, we have built a trap using a gold wire filled PCF cane. 19~wires protrude 50\um from the end of the glass, forming a regular pattern of electrodes. To trap ions we approximate a ring trap by applying radio frequency~(RF) voltages to those electrodes as sketched in figure\,\ref{fig:RingTrap}.

We used the stack-and-draw technique as described in the previous section to produce a fibre cane consisting of three concentric rings of capillaries around a centre capillary. In total this results in a structure of 37 holes ordered in a honeycomb lattice. The distance between the centres of any two holes is 150\um and the hole diameter is 145\um. The overall diameter of the cane is 1.5\,mm. In one run of the drawing process $\sim$10\,meters of cane was produced. For the final device a 10\,mm piece was cleaved off and 30\,mm long 127\um thick gold wires (MaTecK Au Wire 0.127 mm diameter) were manually inserted in the 19 inner holes from one side, the bottom side of the future device, such that they protruded from the top by up to 2\,mm. This process was monitored using an optical microscope.

\begin{figure}
\includegraphics{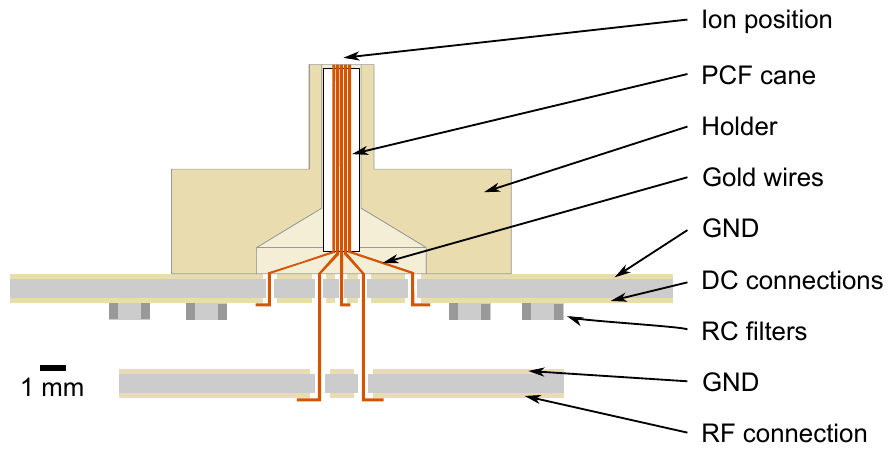}
\caption{
Sketch of the PCF cane trap, its holder and its connection to two circuit boards, one for DC connections, one for RF.
}
\label{fig:trapAndHolder}
\end{figure}

To obtain all electrodes at the same height and to have a smooth surface the wires protruding from the top were polished down to roughly 50\um above the glass surface (using Thorlabs Fiber Polishing/Lapping Film down to 0.3\um grit). For the polishing step the wires at the top were temporarily covered in a resin to keep the soft gold wires in place and prevent gold dust from getting in the gaps between protruding wires. At the other end of the glass structure we encountered the problem that during insertion chips of gold got ripped off by the sharp edges of the glass cane. To avoid electrical shorts the silica cane was re-cleaved 3\,mm away from the bottom. Removing 3\,mm of glass left clean wires at the point where they leave the cane. These wires were fixed to the glass structure at this position using an insulating Epoxy (Epo-Tek 353ND-T). The individual wires were then separated and attached to tracks on a vacuum compatible circuit board (Rogers coper-coated 0.03" RO3003 substrate) that carries low pass RC filters (R = 100\,k$\Omega$, C = 840\,pF). All electrical connections were made using conducting epoxy (Epo-Tek H20E). The final trap assembly is sketched in figure\,\ref{fig:trapAndHolder}.

A limitation to the desirable length of the device is the capacitance between the wires that run down the fibre cane. To measure the reactance of the trap assembly it was connected to a tunable test circuit containing an inductive element and brought to resonate at different frequencies. For each setting of the test circuit the trap assembly was replaced by capacitors until the same resonance frequency was obtained. The capacitive part of the reactance of the trap and circuit board combined was extracted as $31\pm1$\,pF. Simulations show that the wires in the PCF cane should have a capacitance of 13\,pF. This number is proportional to the cane's length but independent to scaling in the radial direction. By approximating the circuit board with a parallel plate capacitor we estimate it to have a capacitance of 15\,pF and attribute the rest to the wires going to the PCF cane.

The trap was placed in an ultra high vacuum~(UHV) chamber and RF is applied to the trap electrodes by forming a resonance circuit between the capacitance of the board and trap and an inductive secondary coil wound on a toroidal core (MICROMETALS, INC. T80-10) which was placed outside the vacuum chamber\,\cite{13Allcock}. The loaded $Q$-value of the resonator was measured to be $64.4$ and a voltage step up of $9.5$ is achieved for inductively in-coupled RF.

\section{Trap operation and characterization}

The electrode structure of the trap is shown in figure\,\ref{fig:RingTrap}. The trap was wired up such that we approximate a surface-electrode ring trap by having the central electrode surrounded by a first ring of 6 electrodes and a second ring of 12 electrodes. We apply RF voltages to the first ring while maintaining all other electrodes at RF ground. This forms a 3-dimensional confining pseudopotential 88\um above the electrode surface~\cite{05Chiaverini2}. By applying static potentials~(DC) to the individual RF ground electrodes, we can compensate electrical stray fields and manipulate the potential.

We trap singly charged Calcium ions by applying RF at 40.3\,MHz with a 70\,V zero-peak amplitude. Calcium atoms are evaporated from a resistive oven and photo-ionized at the trapping site~\cite{01Gulde}. With constant Doppler cooling applied ions stayed trapped for more than five hours. Potential candidates for ion loss are collisions with background gas atoms or molecules. One possible concern of our trap design was that gases trapped in the unfilled capillaries of the PCF cane or in gaps due to the size mismatch of gold wires and capillaries might lead to a virtual leak close to the ion. Assuming pockets of the approximate volume of our unfilled capillaries initially filled with Nitrogen at atmospheric pressure, which are connected by the capillaries to vacuum at $10^{-11}$\,mbar, we can use the ideal gas and Hagan-Poiseville equations\,\cite{98Hansen} to estimate the number of particles in the pockets and thus the resulting leak rate. After one week at room temperature we calculate a residual leak rate of $\sim10^{-17}\,$mbar\,l/s. This is negligible compared to the overall rate in our system which we estimate from the $\sim 10^{-11}$\,mbar pressures and our nominal 100\,l/s pumping speed to be $\sim10^{-9}\,$mbar\,l/s. The depth of our trapping pseudo-potential well was simulated to be 100\,meV.

The DC electrodes can be used to manipulate the trapping potential. The trapping potential~$V$ can be expanded to second order about the null position of the pseudopotential as $V = \sum_i \partial_{q_i}V q_i +\sum_{jk} H_{jk} q_j q_k$ where $q_i, q_j$ and $q_k$ are co-ordinates in three orthogonal spatial directions and the subscript indices run over the co-ordinate axes $x,y,z$ denoted in figure \ref{fig:RingTrap}. The eigenvalues of the Hessian $H_{jk}$ are proportional to the square of the trap frequencies. The chosen co-ordinate system is close to the principal axes, thus small angles of rotation about the principal axes $\theta_i$ can be approximated by changes to off-diagonal terms in $H_{jk}$, and a change in the curvature along one co-ordinate axis approximates a change in the curvature along the aligned principal axis (this is proportional to the trap frequency $\omega_i$ along that axis squared). The gradient $\partial_{q_i}V$ results in displacement $D_i$ of the trapping center along $q_i$. A potential applied to any electrode affects all of these parameters. For ease of operation linear combinations of electrode potentials which uniquely affect one of these parameters were found using simulations based on the boundary element method\,\cite{BEM,96Korsmeyer}.  Three example sets for generating an electric field or changing the trap frequency along $x$ and for approximating a tilt around the $x$ axis are given in table\,\ref{tab:voltages}.

\begin{table*}
\begin{tabular}{l*{13}{c}r}
electrode         & 1 & 2 & 3 & 4 & 5  & 6 & 7& 8 & 9 & 10 & 11 & 12  & 13 \\
\hline
set for $\Delta D_x=$ 1\um [V]       &0.00 & -1.16 & 1.84 & 1.26 & -2.73 & 0.00 & 0.28 & 0.00 & 0.39 & -0.60 & 0.00 & 1.06 & 0.02 \\
set for $\Delta \omega_x =$ 0.1\,MHz [V]    &0.00 & -3.50 & 2.86 & 2.72 & -7.99 & 0.00 & 6.01 & 0.00 & -3.01 & -1.43 & 0.00 & 1.72 & -0.11\\
set for $\Delta \theta_x =$ 0.1\,$^{\circ}$ [V]    & -1.68 & 0.00 & 2.79 & -1.78 & 0.00 & 1.69 & -1.90 & 0.00 & 0.45 & 0.12 & 0.00 & 0.58 & 0.01 \\
final voltages [V]      & 0.53 & 4.14 & -5.55 & -2.08 & 6.84 & -0.36 & -1.13 & 0.00 & -0.81 & 1.49 & 0.00 & -1.96 & 0.01 \\
\end{tabular}
\caption{
Sets of simulated voltages that produce a displacement in $x$-direction by 1\um, a change of the trap frequency by 0.1\,MHz in this direction and a tilt of 0.1\,degrees in the plane perpendicular to this direction when applied to electrodes as specified in figure\,\ref{fig:RingTrap}. Similar sets were produced for $y$ and $z$ directions. The last row shows the working voltages after micromotion compensation. Electrode~11 was shorted to ground due to an error in the fabrication process, but this does not significantly restrict our control.
}
\end{table*}\label{tab:voltages}

In order to Doppler cool an ion, all three principal axes of the ion's motion must have a finite projection onto the k-vector of the cooling laser. In our setup, the laser beams are directed parallel to the trap surface. This could cause problems if the applied static potentials were symmetric about the $z$-axis, since this would result in one of the principal axes being vertical.  In order to avoid this, we tilt the axes of the static potential in the direction of the cooling laser by an angle which we estimate from simulations to be around 2.4\,degrees. We observe that trapping is also possible when all DC electrodes are set to ground, which could be due to imperfections in the position of electrodes or patch potentials on electrodes which result in a residual tilt of the principal axes of the ion's motion.

Electric stray fields can displace the ion from the zero position of the potential generated by the RF electrodes. It then experiences a driven micromotion at the RF frequency\,\cite{98Berkeland2}. This motion can be seen as sidebands if a laser frequency is scanned across a resonance of the ion. The micromotion-sidebands were simultaneously minimized on two perpendicular lasers pointing along $x-y$ and $x+y$. An example set of optimized potentials is shown in the last row of table\,\ref{tab:voltages}.

Over the period of 3\,months drifts in the electric stray fields of $\sim100$\,V/m were observed with the fields apearing stable over the course of days. A possible source of stray fields are charges that get induced on dielectrics exposed to laser light~\cite{10Harlander}. We think that our 50\um high electrodes should help shielding the glass surface from incident light and shield the ion from charge build up on the insulator.

Using a narrow linewidth laser at 729\,nm on the $S_{1/2}-D_{5/2}$ quadrupole transition of Ca$^+$ we are able to resolve the sidebands which result from secular motion of the ion along the principal axes of its confining potential\,\cite{00Roos}. We measure secular oscillation frequencies of 3.8\,MHz, 2.0\,MHz and 1.8\,MHz. These numbers agree with simulations and we conclude that the first is related to motion perpendicular to the electrode surface, while the latter are close to being parallel with the $x$ and $y$ directions respectively.

\begin{figure}
\includegraphics{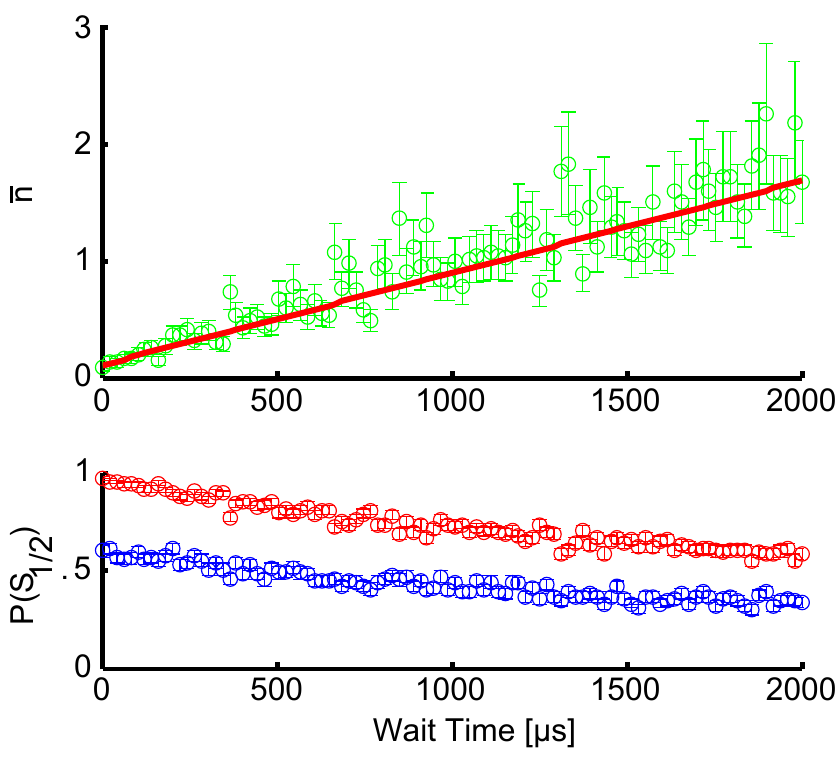}
\caption{
Heating rate measurement. The lower plot shows the probability of the ion to be left in the electronic ground state of the ion ($S_{1/2}$) after driving on a motion adding (lower) or motion subtracting (upper trace) sideband. Every datapoint consists of 300 measurements.\\
In the upper plot the average number of quanta of motion (assuming a thermal state) as calculated from the two sidebands is shown together with a weighted linear fit to the data. The slope of this fit gives a heating rate of 787$\pm$24\,quanta/s.
}
\label{fig:nBar}
\end{figure}

Following Doppler cooling, we sideband cool the 1.8~MHz mode close to the ground state of motion using the $S_{1/2}-D_{5/2}$ transition and dipole-allowed $D-P$ transitions at 854\,nm and 866\,nm. We probe the temperature of the mode after sideband cooling using sideband thermometry \cite{95Monroe,00Turchette}, and deduce a final mean occupation of $\overline{n} = 0.08\pm0.03$\,quanta. We also probe the mean excitation as a function of a delay introduced between the sideband cooling and the probe pulse, resulting in the data shown in figure\,\ref{fig:nBar}. The deduced heating rate is 787$\pm$24\,quanta/s. This allows us to estimate a spectral noise density at the ion of $S_E = 9.7\pm0.3\times 10^{-12}$\,(V/m)$^2$/Hz. This is at the lower range of reported heating rates for room temperature traps operated under similar conditions~\cite{14Daniilidis}. We note that no particular care was taken in order to make sure that the surface was clean, however no photo-lithography has been performed. Baking of the vacuum system for a week at 160\,$^{\circ}$C was performed on two separate occasions with the trap present.

\section{Conclusion}
We have presented the operation of a 100 micron scale ion trap built from a photonic crystal fibre cane. Once the cane had been produced, no specialist techniques such as photo-lithography were used. In contrary to most other ion traps in use today the electrical connections go straight to the back of the device, leaving an open geometry which may be of use for laser beam access, integration with optical cavities \cite{13Brandstaetter,13Steiner}, or for precision measurements of surfaces \cite{09Maiwald}. Optical cavities could be useful for generating high light intensities at the ion for optical trapping \cite{10Schneider}.

In future, we anticipate the possibility to integrate PCF light guiding technologies (or other optical fibres) into traps similar to that described here. This could be realized by replacing one of the gold wires with a PCF, although this would require a change in the electrode geometry or the applied potentials. Further minitiurization of the current trap, using the full toolkit of gold-filled PCF production, might allow ion trapping at the tip of an optically guiding PCF, and provide a route towards small two dimensional arrays of ion traps connected by optical fibre links with inherent fibre optic access to single trapping sites.

\begin{acknowledgments}

The authors thank Robert Davis for initial simulations and Fabian Fuchs for assisting in the fibre drawing process and initial work on the gold wire filling. We would further express gratitude to H.-Y. Lo, F. Leupold, L. de Clercq, V. Negnevitsky, M. Marinelli and U. Sol\`{e}r for their work on electronics, experimental control and laser-system used.
We thank the NIST group of D. Wineland for sharing their Boundary Element Method simulation code.
We acknowledge support from the Swiss National Science Foundation under grant number 200021 134776, through the National Centre of Competence in Research for
Quantum Science and Technology (QSIT) and Erlangen Graduate School in Advanced Optical Technologies (SAOT).

\end{acknowledgments}


%
%

%


\bibliography{myrefs}

\end{document}